\documentclass[acmlarge]{acmart} 
\usepackage[utf8]{inputenc}
\usepackage[T1]{fontenc}
\usepackage{url}
\usepackage{soul}
\usepackage{algorithmic}
\usepackage{graphicx}
\usepackage{tabularx}
\usepackage{longtable}
\usepackage{xltabular}
\usepackage{makecell}
\usepackage{enumitem}
\usepackage{textcomp}
\usepackage{xcolor}
\usepackage{colortbl}
\usepackage{gensymb}
\usepackage{multirow}
\usepackage{subfig}
\usepackage{hyperref}
\usepackage{array}
\usepackage{subfig}
\usepackage{hyperref}
\usepackage{xcolor}
\usepackage[makeroom]{cancel}

\definecolor{susu}{rgb}{0, 0.843, 0.161}
\definecolor{susd}{rgb}{0.965, 0.718, 0}

\AtBeginDocument{%
  \providecommand\BibTeX{{%
    \normalfont B\kern-0.5em{\scshape i\kern-0.25em b}\kern-0.8em\TeX}}}

\makeatletter
\newcommand{\Cut}[1]{}
\newcommand{\thickhline}{
    \noalign {\ifnum 0=`}\fi \hrule height 1pt
    \futurelet \reserved@a \@xhline
}

\newcolumntype{"}{@{\hskip\tabcolsep\vrule width 1pt\hskip\tabcolsep}}
\makeatother

\setcopyright{acmcopyright}
\copyrightyear{2024}
\acmYear{2024}
\acmDOI{XXXXXXX.XXXXXXX}

\newcommand{\anonymized}[1]{{\textcolor{orange}{\MakeUppercase{ANONYMIZED}}}}
\newcommand{\system}{\emph{MapIO}}

\acmJournal{IMWUT}

\begin{document}
\title{MapIO: Embodied Interaction for the Accessibility of Tactile Maps Through Augmented Touch Exploration and Conversation}

\author{Matteo Manzoni}
\email{matteo.manzoni2@studenti.unimi.it}
\affiliation{%
 \institution{University of Milan}
 \streetaddress{Via Giovanni Celoria, 18}
 \city{Milan}
 \state{}
 \country{IT}
 \postcode{20133}
}

\author{Sergio Mascetti}
\email{sergio.mascetti@unimi.it}
\affiliation{%
 \institution{University of Milan}
 \streetaddress{Via Giovanni Celoria, 18}
 \city{Milan}
 \state{}
 \country{IT}
 \postcode{20133}
}

\author{Dragan Ahmetovic}
\email{dragan.ahmetovic@unimi.it}
\affiliation{%
 \institution{University of Milan}
 \streetaddress{Via Giovanni Celoria, 18}
 \city{Milan}
 \state{}
 \country{IT}
 \postcode{20133}
}

\author{Ryan Crabb}
\email{ryan.crabb@ski.org}
\affiliation{%
 \institution{The Smith-Kettlewell Eye Research Institute}
 \streetaddress{Fillmore St., 2318}
 \city{San Francisco}
 \state{CA}
 \country{USA}
 \postcode{94115-1813}
}

\author{James M. Coughlan}
\email{coughlan@ski.org}
\affiliation{%
 \institution{The Smith-Kettlewell Eye Research Institute}
 \streetaddress{Fillmore St., 2318}
 \city{San Francisco}
 \state{CA}
 \country{USA}
 \postcode{94115-1813}
}

\begin{abstract} 
For individuals who are blind or have low vision, tactile maps provide essential spatial information but are limited in the amount of data they can convey. Digitally augmented tactile maps enhance these capabilities with audio feedback, thereby combining the tactile feedback provided by the map with an audio description of the touched elements.
In this context, we explore an embodied interaction paradigm to augment tactile maps with conversational interaction based on Large Language Models, thus enabling users to obtain answers to arbitrary questions regarding the map.
We analyze the type of questions the users are interested in asking, engineer the Large Language Model's prompt to provide reliable answers, and study the resulting system with a set of 10 participants, evaluating how the users interact with the system, its usability, and user experience.
\end{abstract}



\maketitle

\section{Introduction}
\label{sec:intro}
Tactile Maps (TMs) are essential tools for Blind and Low Vision (BLV) people, providing spatial information through raised lines, symbols, and braille. However, these maps can convey limited amount of information due to space constraints and the nature of tactile encoding. Digitally Augmented Tactile Maps (DATMs) have emerged as a solution to this limitation by integrating digital technologies that offer additional, dynamic information.
DATMs typically provide audio feedback related to specific Points of Interest (POIs) that the user is exploring, enhancing the amount and richness of spatial and contextual information available.

Despite the advantages of DATMs in enhancing tactile information with additional feedback, a significant restriction remains: they are specifically designed to address a predefined set of problems, which limits their adaptability to different user needs or contexts. Adding new features or extending their capabilities requires modifications to the underlying code. This rigidity poses challenges for customizing DATMs to support diverse tasks or providing tailored experiences for users, ultimately dampening the potential of these maps to be versatile, user-centric tools.

To address the limitations of current DATMs, this paper proposes a novel embodied interaction paradigm using tactile exploration and conversational querying.
The approach, implemented as a DATM, allows users to verbally ask questions, providing dynamic and context-sensitive responses.
The system, called \system{}, leverages Large Language Models (LLMs) to provide answers to a wide range of questions without requiring these specific functionalities to be implemented.
The LLM's response is then read aloud to the user.

The methodology follows a user-centric, iterative design approach structured in five phases.
First, through a formative study conducted with $5$ BLV participants and Teachers of BLV students (TBLV) we identify the goals, tasks, and types of questions that BLV people have when using TMs.
In particular, we collect a set of benchmark questions that the participants would ask to a DATM augmented with conversational interaction capabilities.
Second, based on the outcomes of the formative study, we design a system, \system{}, that integrates a LLM and tracks the user's fingers while exploring a TM using computer vision.
\system{} enhances the LLM's ability to respond accurately by supplementing the user queries with additional context, including the point on the map the user is touching, relevant map details not inherently known to the LLM (\textit{e.g.}, POI on the map), and guidelines provided to the LLM to improve the answer quality and reduce the risk of hallucinations.
We iteratively refine the prompt augmentations until satisfactory results are achieved, measured as the truthfulness and completeness of the LLM's responses to the benchmark questions.
Third, we conduct a preliminary observational study with $2$ participants during which the conversational interaction is simulated by a supervisor who provides answers by reading LLM responses, correcting obvious errors.
Fourth, based on participants' interactions and the types of queries they ask, we update the system design and prompt augmentations accordingly, and we conduct a second preliminary study with $3$ participants, testing the modifications.
Fifth, we conduct a user study with $10$ participants, using the final version of the system, focusing on how participants interact with the system and how they manage errors,
incomplete information, and uncertainty during their interactions.
We also study system usability.

We address the following research questions, considering a system that enhances TMs with the possibility to ask arbitrary questions related to a position pointed on the map:
\begin{itemize}
\item[\textbf{Rq1}] What kinds of questions would BLV people ask?
\item[\textbf{Rq2}] Can these questions be reliably answered by an existing LLM model?
\begin{itemize}
\item[\textbf{Rq2.1}] How to design LLM \textit{query context} to augment the user prompts in order to achieve accurate answers?
\end{itemize}
\item[\textbf{Rq3}] How would the user interact with the TM and such a system?
\item[\textbf{Rq4}] How would users perceive such a system in terms of usefulness, usability, workload, and overall user experience?
\end{itemize}

The results show that users are interested in learning about POIs, their properties (\textit{e.g.}, opening hours), road accessibility, and directions. While existing LLMs have limitations in spatial reasoning \cite{openaiLimitations}, prompt engineering can enhance their performance to near-perfect results. Effective prompts should incorporate a structured representation of the road network, including nodes, edges, and accessibility data (\textit{e.g.}, type of paving) alongside POI details like location and operating hours. 


\section{Related Work}
\label{sec:related}

Tactile materials are indispensable tools for BLV people to access information \cite{zebehazy2014straight}.
In particular, TMs are used in contexts such as Orientation \& Mobility training~\cite{wright2010best, toyoda2020effects, ungar1993role} or navigation in unfamiliar environments~\cite{ungar1994can, toyoda2020effects, ungar1993role}.
Braille is sometimes used to label key features, but the text is often abbreviated, requiring a separate braille keymap~\cite{jimenez2009biography}.
Additionally, many BLV people do not read braille~\cite{wiazowski2014can}.

Previous literature has addressed the problem of digitally improving the accessibility of tactile materials, such as embossed paper drawings or 3D printed models. This body of research is described in Section~\ref{ssec:datm}.
Our work contributes to this research domain through the design of a novel embodied interaction paradigm for the accessibility of TMs that combines manual exploration and conversational querying using LLMs.
To this end, we also survey prior research in the field of assistive technologies based on the use of conversational interfaces,
in particular for the accessibility of tactile materials (see Section~\ref{ssec:llm_assistive_technologies}).

\subsection{Digitally Augmented Tactile Supports}
\label{ssec:datm}

Augmenting tactile materials with audio feedback can improve their accessibility \cite{brock2015interactivity}.
For instance, the T3 Tactile Tablet \cite{t3TactileTable} from Touch Graphics allows users to overlay a tactile graphic on an Android tablet, enabling the device to react to touch by providing audio feedback for each tactile feature on the graphic.
Such systems are typically limited to 2D tactile graphics that fit on a tabletop touchscreen, thin enough to allow the touchscreen to sense the locations of fingertips on the tactile graphic.

To extend accessibility to 3D tactile supports, computer vision techniques can be employed using RGBD cameras \cite{shen2013camio, wang2023touchpilot}, standard webcams \cite{shi2017markit}, and smartphone applications \cite{coughlan2020towards, fusco2015tactile, zeinullin2022tactile}.
These methods must reliably interpret user intent.
For example, some systems require a specific pointing tool \cite{coughlan2020towards}, while others ask users to augment their fingertip with a visual marker \cite{shi2017markit}.
These methods, however, are often perceived as unnatural \cite{coughlan2020towards, coughlan2022non} and may not support simultaneous interactions with multiple hands or fingers.
With the rise of advanced hand-tracking algorithms, touch-based interfaces are becoming more natural, intuitive and versatile \cite{wang2023touchpilot, zeinullin2022tactile, narcisi2024accessible}.

DATMs provide additional information regarding the TM and help users locate features of interest \cite{gotzelmann2016lucentmaps}.
For example, Wang et al. \cite{wang2009instant} proposed a system that converts an image of a map into a tactile format, which can then be printed using a tactile printer. This TM can be placed on a touchpad, allowing users to explore it while the system provides audio feedback about the touched region. 
This approach is similar to the one explored in \cite{miele2006talking}, with the main difference being that the latter uses the T3 tablet instead of a touchpad.

Despite these advances, a major limitation remains: the way users access information is predetermined.
For instance, in \cite{coughlan2020towards}, users point to a feature on a 3D map using a stylus and receive the name of the location but cannot obtain additional details.
In \cite{narcisi2024accessible} and \cite{miele2006talking}, users can receive varying levels of detail based on gesture type, but the information is still static, and users cannot ask for clarifications or further details.
Our research addresses this limitation by integrating a TM with an LLM, enabling users to ask arbitrary questions about it.

\subsection{Use of conversational interfaces in Assistive Technologies}
\label{ssec:llm_assistive_technologies}

The main limitation of the projects discussed in the previous section is that users interact with the map solely through touch, which restricts the amount of information that can be conveyed. This issue can be mitigated by introducing a conversational interface, allowing users to directly ask the system questions and receive answers.
An example of a conversational interface applied to tactile maps (TMs) is LucentMaps \cite{gotzelmann2016lucentmaps}, a mobile app that enables interaction with 3D portable TMs through touch and speech. Similarly, the Jido project \cite{cavazos2019jido} supports interaction with fixed 2D TMs and adds real-world guidance to specific POIs using a companion app.
%
Despite the benefits of these conversational interfaces, both Jido and LucentMaps support only a limited set of queries, which restricts their capacity for comprehensive interaction. Examples of supported queries in Jido include ``Is there a <POI> around?'', ``What is this?'', and ``Guide me to <POI>.'' A more robust conversational interface could be achieved by leveraging LLMs, allowing users to ask a broader range of questions.
%
For example, EasyAsk is an in-app contextual tutorial search assistant designed to simplify app usage on smartphones through voice and touch inputs~\cite{gao2024easyask}. This system leverages LLMs to process spoken requests and contextual information, enhancing accessibility for older adults.
ChartLama \cite{han2023chartllama} and VizAbility \cite{gorniak2024vizability} are LLM-based systems that enhance chart accessibility by allowing users to ask questions about chart content. However, ChartLama is limited to charts generated through a specific pipeline, whereas VizAbility can autonomously interpret chart content using Olli's tree view \cite{blanco2022olli}, making it easier to integrate into existing screen readers.
Similarly, Cuadra et al. \cite{cuadra2024digital} used an LLM-based system to improve accessibility in ``health data entry'' forms. Their system combines LLMs and multimodal interfaces to allow users, including those with disabilities, to complete complex forms independently.

In the navigation field, NaviGPT \cite{zhang2024enhancing} is a mobile application designed to enhance travel experiences for BLV individuals.
Combining obstacle detection, vibration feedback, and an LLM-based conversational interface, NaviGPT provides continuous feedback through image recognition and contextual navigation guidance.
The LLM processes nearby landmarks, routes, and the user's current position, as provided by Apple Maps, to answer user queries. 
Finally, Tran et al. \cite{tran2024enabling} utilize an LLM to support BLV individuals who are learning to use tactile maps. A key distinction between Tran's system and \system{} is that Tran's system operates solely with digital maps of very simple indoor environments, while \system{} aims to support more complex outdoor environments. Additionally, the LLM in Tran's system can answer user queries but lacks knowledge of the user's position on the map and cannot assist with navigation tasks.






\section{Interaction and System Design}
\label{sec:design}
To understand how BLV people use tactile materials (\textbf{Rq3}), and in particular what questions they would ask to a DATM integrated with conversational interaction capabilities (\textbf{Rq1}) we design the the \system{} system and its interaction paradigm through an iterative process structured in 5 steps.
In this section we consider the first four steps, 
while the fifth step, the user evaluation of the final system, is described separately, in Section~\ref{sec:evaluation}.

Section~\ref{ssec:formativeQuestionnaire} describes the formative study used to comprehend how BLV people could use an ideal DATM integrated with conversational interaction capabilities, and to collect questions that users would ask to such a system.
In Section~\ref{ssec:initialPrototype}, we describe the initial design of a novel DATM system based on the outcomes of the formative questionnaire.
Section~\ref{ssec:firstPreliminaryStudy} describes the first preliminary study, used to refine the initial interaction and system design.
Finally, Section~\ref{ssec:secondPreliminaryStudy} describes the second preliminary study, after which we have improved both the interaction design and the prompt augmentations, finalizing the system.

\subsection{Formative questionnaire}
\label{ssec:formativeQuestionnaire}
We conducted a remote study with BLV participants and TBLVs.
Specifically, we designed a questionnaire in which we describe an ideal DATM that is capable of tracking the point of the map touched by the user, reading the names of streets and intersections, and that can answer general questions asked verbally, which can also refer to the current position pointed by the user on the map.
Through two open ended questions, we ask the participants to illustrate what questions they would ask the system and for what purposes they would use it.
The complete questionnaire is reported in Supplemental Material.

The questionnaire was administered via email to four BLV adults. 
Specifically, \textbf{P1}, \textbf{P2} and \textbf{P3} are blind individuals, while \textbf{P4} has low vision.
Three of the them are also TBLV (\textbf{P1}, \textbf{P3}, and \textbf{P4}).
One additional participant, \textbf{P5}, is not BLV and has experience as TBLV.
Table \ref{tab:formativeQuestionnaire-demographic} summarizes participants' demographic data.

\begin{table}[ht]
\centering
\caption{Demographic information on formative questionnaire participants.}
\begin{tabular}{|c|c|c|c|c|c|c|c|}
\hline
\multirow{2}{*}{\textbf{PID}} & \multirow{2}{*}{\textbf{Age}} & \multirow{2}{*}{\textbf{Gender}} & \multicolumn{3}{c|}{\textbf{Visual Impairment}} & \multirow{2}{*}{\textbf{TBLV}} & \textbf{Experience} \\
\cline{4-6}
& & & \textbf{Description} & \textbf{Onset} & \textbf{Residual} & & \textbf{with TMs} \\ \hline
\textbf{P1} & 46 & M & Blindness & birth & No & Yes & High \\ \hline
\textbf{P2} & 46 & F & Blindness & N/A & No & No & Mid \\ \hline
\textbf{P3} & 29 & F & Low vision & birth & Partial & Yes & Mid \\ \hline
\textbf{P4} & 43 & M & Blindness & birth & No & Yes & High \\ \hline
\textbf{P5} & 60 & F & \multicolumn{3}{c|}{None} & Yes & High \\ \hline
\end{tabular}
\label{tab:formativeQuestionnaire-demographic}
\end{table}

All participants expressed an interest in receiving general information about the map, with requests such as \textit{``Give me an overall description of the map''} (\textbf{P1}) or \textit{``What can I find on the map?''} (\textbf{P3}).
Note that these requests would not require to refer to a specific location on the map.
A similar question is \textit{``How many streets are there?''} (\textbf{P2}).
The participants also suggested questions that are related to a specific position on the map.
Specifically, \textbf{P1}, \textbf{P2}, \textbf{P3}, and \textbf{P4} expressed interest in knowing the position they are currently pointing on the map, while \textbf{P1}, \textbf{P2}, \textbf{P3}, and \textbf{P5} also wanted a detailed description of the location.
For example, \textbf{P1} was interested in knowing what they could find along the road that they are touching on the map, while \textbf{P2} expressed interest for more general details about streets.
Accessibility information was mentioned by both \textbf{P4} and \textbf{P5}, with requests like \textit{``Tell me if there are stairs at some point on the road''} (\textbf{P4}).

All participants suggested using \system{} to familiarize themselves with an unfamiliar environment and explore its road network and POIs before actually visiting.
\textbf{P1} and \textbf{P2} also indicated they would use the system to find specific locations on the map, such as \textit{``a street, a school, and so on''} (\textbf{P2}).
Additionally, \textbf{P1}, \textbf{P2}, and \textbf{P5} mentioned that they would use the system to get directions to reach certain places.
\textbf{P1} noted they would ask \textit{``How can I come here from ...?''}, while \textbf{P4} mentioned they would use the system \textit{``to explore a route and decide whether I can go on foot or by taxi or other public transportation.''}.
\textbf{P5} also stated they would use the system to teach BLV individuals how \textit{``to understand a complex intersection and/or traffic patterns, route planning, etc.''}.

Ultimately, based on the participants' answers, we curated a list of benchmark questions (\textbf{Rq1}) that our system should be capable of addressing (reported in Supplemental Material).
The benchmark questions were used to design the LLM prompt, described in Section~\ref{sec:engineering}.

\subsection{Initial system and interaction design}
\label{ssec:initialPrototype}
Motivated by the limitations with the existing DATM approaches (see Section~\ref{ssec:datm}), and the use cases reported by the participants of the formative study (see Section~\ref{ssec:formativeQuestionnaire}), we designed a novel DATM system, called \system{}.
The system is designed to enable embodied interaction with a TM, augmenting manual exploration with speech feedback related to the touched elements, and supporting conversational querying that also considers contextual information related to the map and the knowledge of the elements touched by the user.
The system design is described in Section~\ref{sssec:app} while the interaction design is described in Section~\ref{sssec:interaction}.


\begin{figure}
    \centering
    \includegraphics[width=0.55
    \linewidth]{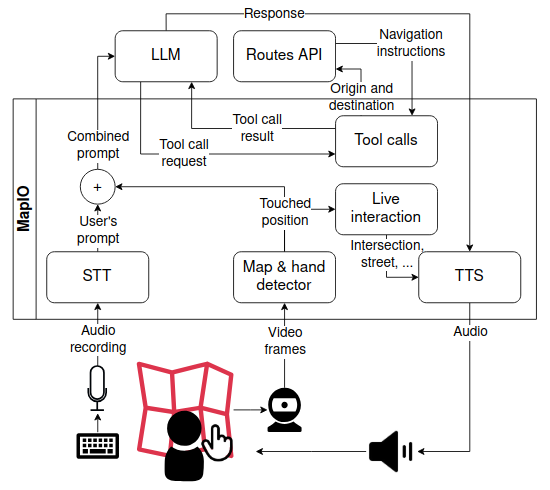}
    \caption{\system{} initial design architecture}
    \Description[Architecture of the initial design of \system{}]{A user interacts with a tactile map and points a location on it. A camera is placed above the maps and sends video frames to the map \& hand detector module of \system{}. This computes the touched position on the map, which is processed by the live interaction module to get a description of the position. The description is converted to audio through TTS and played back to the user.
    A keyboard is placed on the leaf of the map: when the user presses the space bar the microphone is used to record their verbal request. The request is converted into text through STT and treated as the user's prompt. This is combined with the touched position produced by the map \& hand detector module into the combined prompt and sent to the LLM. The LLM generates a response, which is converted into audio though TTS and played back to the user. To generate a response the LLM can request to \system{} to execute local tool calls. The navigation tool call forwards the request to Google Routes API. The LLM and Routes API modules are rendered separately from the other modules as they are not part of the \system{} software.}
    \label{fig:architecture}
\end{figure}

\subsubsection{System Design}
\label{sssec:app}
The \system{} is implemented as a computer application within the architecture depicted in Fig.~\ref{fig:architecture}.
A video camera is used to acquire a video stream of a TM placed on a desk, and of the hands of a user interacting with the map, at approximately $10$ frames per second.
The images are processed by the \textit{Map \& hand detector} module, that detects the map by matching SIFT features \cite{zhou2009object} of a map image template (previously stored by \system{}), and those extracted from video frames.
The same module computes a homography matrix to convert the coordinates from the frame reference system to that of the map.
Finally, the \textit{Map \& hand detector} tracks the hands and checks if there is a pointing gesture.
This is achieved by using the MediaPipe library \cite{zhang2020mediapipe} that produces a list of $20$ landmarks for each hand.
If the landmarks of the index finger are all aligned, and those of the other fingers are not, a pointing gesture is detected (thumb landmarks are ignored).
The pointing position corresponds to the landmark of the tip of the index finger, converted to the map reference system.
To reduce tracking noise, the detected index positions are averaged over a $10$-frames window.

To enable audio feedback when elements of the map are touched, \system{} also has a map model
containing a graph of the streets and intersections displayed on the map, POIs present in the area, and additional information about those elements, such as opening hours, facilities, and accessibility features or barriers.
Through computer speakers, the \textit{Live interaction} module provides audio feedback on the elements touched by the user.
These can be sound cues, or verbal messages synthesized using a Text-to-Speech (TTS) engine.

Upon pressing the keyboard space bar key, the user can verbally ask queries that are recorded using the microphone, and converted into text by the Speech-to-Text (STT) module\footnote{This module relies on Google Speech-to-Text (STT) API \url{https://cloud.google.com/speech-to-text}.}.
The text transcription is then forwarded to a LLM\footnote{We used the ChatGPT-4o-2024-08-06 model, accessed through its API.}, along with the touched position and relevant information about the map model, as described in Section~\ref{sec:engineering}.
\system{} also leverages the capability of the adopted LLM model to call local functions - called \textit{tool calls} - for completing specific tasks.
Each tool call includes a description of the provided functionality and the explanation of when the LLM should call it.
When the LLM decides to use a tool call it returns a specific request that contains the name of the function to execute and its parameters.
\system{} then executes the function and sends the result back to the LLM, which ultimately produces the final answer to the user's query.
In particular \system{} uses tool calls to intercept navigation requests from the user and forward them to Google Routes API\footnote{\url{https://developers.google.com/maps/documentation/routes}} that, given a starting point and a destination, would return street-by-street indications on how to navigate between the two.
This approach was required to address the poor LLM's spatial reasoning capabilities~\cite{openaiLimitations}.
The answer provided by the LLM is played using the TTS.

Note that, when the user points to something on the map, the feedback is provided almost instantaneously. Instead, the total time to compute an answer to a vocal query can vary from a fraction of a second to more than ten seconds. 
During this time, the system does not provide audio feedback when the user is pointing elements on the TM.
Instead, every $7$ seconds, a message informs the user that the system is computing the answer.

\subsubsection{Interaction Design}
\label{sssec:interaction}
We aimed to provide a natural interaction paradigm, based on manual and conversational interaction.
The manual interaction is similar to prior approaches~\cite{brock2015interactivity,coughlan2020towards,coughlan2022non,fusco2015tactile,shen2013camio,shi2017markit}, in particular those using computer vision-based hand tracking~\cite{wang2023touchpilot,zeinullin2022tactile,narcisi2024accessible}, and requires the user to point tactile elements on the map with the index finger.
When using the pointing functionality, the system will read the name of the touched element, and the reading stops if the finger moves away from the element.
When no pointing hand is found, the system plays a background sound of crickets chirping.
Thus, the layout of tactile materials can be freely explored with open hands without triggering audio feedback.

The user can also verbally ask information regarding streets, intersections, and POIs present on the map.
For streets, this information includes the paving material, slope (if any), traffic direction (if one way street), and accessibility features or barriers (\textit{e.g.}, presence of tactile pavings).
For intersections, it includes the type of the intersection (\textit{e.g.}, a ``T'' intersection), presence of crosswalks or traffic lights, and accessibility features (\textit{e.g.}, crosswalk audio signal).
For POI, the features may include opening times, facilities (\textit{e.g.}, WiFi) and accessibility features (\textit{e.g.}, presence of a TM of the venue).

To ask questions, the user points an element on the map and presses the space bar on the keyboard.
When the space bar is pressed again, the recording of the question stops and the captured audio is converted into text using STT.
A notification sound indicates the start and the end of registration, and a chime sound is played at the end of the returned message.
%
%
%
The system supports additional interactions through the computer keyboard.
Users can stop the current system announcement by pressing the \textit{escape} key, while the \textit{enter} key pauses and resumes the announcements.

\subsection{First preliminary study}
\label{ssec:firstPreliminaryStudy}
We evaluated the initial system and interaction design through a preliminary observational study conducted with $2$ participants, one blind (\textbf{P6}) and one with low vision (\textbf{P7}) (see Table~\ref{tab:fps-demographic}).
Both had prior experience as TBLV, and both are proficient with braille and TMs.
They have also previously participated in studies using digitally augmented tactile supports.

\begin{table}[ht]
\centering
\setlength{\tabcolsep}{0.2em}
\caption{Participants in the first preliminary study. LP: light perception, DATS: digitally augmented tactile supports}
\begin{tabular}{|c"c|c|c|c|c|c|c|c|c|}
\hline
\multirow{2}{*}{\textbf{PID}} & \multirow{2}{*}{\textbf{Age}} & \multirow{2}{*}{\textbf{Gender}} & \multicolumn{3}{c|}{\textbf{Visual Impairment}} & \multirow{2}{*}{\textbf{TBLV}} & \multicolumn{2}{c|}{\textbf{Expertise}} & \textbf{Experience} \\
\cline{4-6}
\cline{8-9}
& & & \textbf{Description} & \textbf{Onset} & \textbf{Residual} & & \textbf{Braille} & \textbf{TMs} & \textbf{with DATS} \\
\thickhline
\textbf{P6} & 47 & F & Fully blind & birth & No & Yes & High & High & Yes \\ \hline
\textbf{P7} & 37 & M & LCA~\cite{den2008leber} & birth & LP & Yes & Mid & Low & Yes \\ \hline
\end{tabular}
\label{tab:fps-demographic}
\end{table}

\subsubsection{Study design}
We used the same experimental methodology as the main evaluation~\ref{sec:evaluation}, with the following  differences.
In the preliminary studies, laptop space bar and escape keys on the keyboard were used instead of external buttons.
The system implemented the Iteration 6 of the prompt augmentations (see Section~\ref{sssec:iterations6and7}).
Since we were still working on refining the LLM component to get reliable answers, we deactivated the TTS reading of the LLM output.
Instead, an experimenter would read the text answers provided by the LLM, correcting them if needed (similar to a Wizard-of-Oz approach).
However, as no major corrections were needed, we enabled direct conversational interaction in the second preliminary study (see Section~\ref{ssec:secondPreliminaryStudy}).
Due to the limited amount of data, we directly report our observations instead of performing thematic analysis.

\subsubsection{Results}
Both participants used both hands to explore the map, often pointing with one hand and exploring with the other.
Sometimes, a pointing hand outside of the map area was erroneously tracked.
Being proficient with TMs and braille, both relied mostly on their tactile exploration abilities and audio feedback from the system instead of asking directions to the LLM.
The audio feedback seemed invaluable to disambiguate situations where braille labels on the map border were in the vicinity of multiple streets.
However, we also noticed situations in which this strategy was not sufficient.
For example, \textbf{P6} could not find a POI on a short street in the middle of the map that, due to space constraints, did not have a braille label.
Querying the system provided immediate guidance to reach the target POI.

When asking for guidance, both participants seemed to have difficulties in memorizing instructions, in particular those with multiple steps.
Thus, the participants would coarsely explore the area referring to whatever they remembered of the instructions.
\textbf{P6} also highlighted that there was no way to know if they have reached the destination if not by asking the system.
The instructions were perceived to be overly verbose as well, since they contained also accessibility instructions regarding each traversed street and intersection.
\textbf{P7} also asked for distance in the guidance steps, and \textbf{P6} queried the system for the same information.

\subsubsection{Outcomes}
To address these comments and observations, we applied the following modifications.
First, we removed the tracking of hands outside the map area to avoid unintended interactions.
Second, we added a new tool call (see Section~\ref{sssec:app}) in order to have guidance instructions given one at a time.
The user would then need to ask the next instruction to the system once the previous one is completed (\textit{e.g.}, once a turning point is reached).
We also added another tool call that activates sound feedback also for POI relevant to user's last question, making it possible to explore them on the map
(see Section~\ref{sssec:iterations6and7}).
We further added the distance information to every navigation step, and we reduced the verbosity of the guidance instructions by removing accessibility indications.
These could still be accessed by explicitly asking them to the LLM.
Additionally, we increased the frequency of the position verbal messages, improved the accuracy of the pointing position localization by doubling the size of the window on which the average is computed, corrected minor bugs with the street graphs of our maps, and added ``gravity'' to streets, nodes, and POI, so that small inadvertent hand movements from the user do not result in leaving the currently indicated feature.

\subsection{Second Preliminary Study}
\label{ssec:secondPreliminaryStudy}
The second preliminary study was conducted with 3 participants.
One of them is blind (\textbf{P8}) and two have some residual vision (\textbf{P9}, \textbf{P10}).
\textbf{P9} also has reduced touch sensitivity, which impacted some of the study tasks.
Indeed the participant did not complete most of the study tasks.
Participants had various levels of expertise with braille and LLMs, but all had relatively low expertise with TMs.
In particular \textbf{P9} had prior experience as TBLV, and was the only participant who had experience with digitally augmented tactile supports, despite a low expertise in TMs (see Table~\ref{tab:fps-demographic}).

\begin{table}[ht]
\centering
\setlength{\tabcolsep}{0.2em}
\caption{Participants in the second preliminary study. DATS: digitally augmented tactile supports}
\begin{tabular}{|c"c|c|c|c|c|c|c|c|c|c|}
\hline
\multirow{2}{*}{\textbf{PID}} & \multirow{2}{*}{\textbf{Age}} & \multirow{2}{*}{\textbf{Gender}} & \multicolumn{3}{c|}{\textbf{Visual Impairment}} & \multirow{2}{*}{\textbf{TBLV}} & \multicolumn{3}{c|}{\textbf{Expertise}} & \textbf{Experience} \\
\cline{4-6}
\cline{8-10}
& & & \textbf{Description} & \textbf{Onset} & \textbf{Residual} & & \textbf{Braille} & \textbf{TMs} & \textbf{LLMs} & \textbf{with DATS} \\
\thickhline
\textbf{P8} & 75 & F & Fully blind & birth & No & No & High & Low & High & No \\ \hline
\textbf{P9} & 51 & M & CVI~\cite{roman2007cortical} & birth & 20/400 & Yes & None & Low & Mid & Yes \\ \hline
\textbf{P10} & 46 & M & Optic Atrophy~\cite{lenaers2012dominant} & birth & 20/400 & No & Mid & Mid & Low & No \\ \hline
\end{tabular}
\label{tab:sps-demographic}
\end{table}

\subsubsection{Study Design}
The version of the prompt augmentations used in this study was iteration 7 (see Section~\ref{sssec:iterations6and7}).
In this experiment, we used the same methodology as in the previous one.
The only difference for the questionnaire was the addition of
one question asking what the participant would use this system for.

\subsubsection{Results}

All participants had difficulties with the voice recording interaction.
Very often, \textbf{P8} and \textbf{P10} would touch the laptop's touchpad while interacting with the space bar.
Instead, \textbf{P9} would press the touchpad instead of the space bar, and had difficulties in pressing the space bar to start recording and press it again to stop.
Instead, the participant would often talk without pressing the space bar, or would press and hold it while talking (a push-to-talk approach~\cite{woodruff2003push}).
This participant would also often remove hands from the map while asking questions, which resulted in errors when asking questions related to the pointed area.

Pointing interaction felt natural for the participants, but some had differences with the expected interaction.
Specifically, \textbf{P9} would keep the hand raised from the map, pointing slightly from above, which was sometimes harder for the system to track.
Instead, \textbf{P8} would follow the tactile cues with the side of the finger, causing the tracked point to always be a little on the side of the actual tactile element.
This participant slightly moved the finger over the tactile cues continuously, triggering repetitive truncated messages due to the movement: \textit{``I'm not moving, why is it talking to me?''}.
\textbf{P8} also moved quickly over the intersections and wondered why the system did not completely read the intersections names:
\textit{``what is the way to make it say all the intersections?''}
After the explanation, \textbf{P8} learned to keep the pointing stable on tactile cues to have audio feedback read completely.

All participants had difficulties in following the navigation instructions.
In particular, as the turn directions were provided with respect to the route, all participants were confused when their orientation and route orientation differed.
For example, when following a route going towards the bottom of the map, a ``turn left'' instruction means that users should move their finger to the right side of the map.
Distance indications were also not useful to the participants as they would perform large movements even when the instruction mentioned very short distances (in particular \textbf{P8}).
While navigating, \textbf{P8} and \textbf{P9} would not listen to street and intersection names so they would often miss the turning point mentioned in the instruction, which resulted in faults during guidance.
\textbf{P9} also asked whether the chime sound (played at the end of the LLM message) meant that the target point has been reached, suggesting that such a cue would be desirable.
The difficulty in using the navigation functionality elicited alternative strategies from the participants.
Specifically, \textbf{P9} and \textbf{P10} would sometimes try to reach a POI by traversing the streets on the map, searching for the ones in its address.
\textbf{P10} also used a form of ``beaconing'', iteratively moving and asking for the distance to a POI.


\subsubsection{Outcomes}
We addressed the identified issues by implementing the following modifications to the system interaction design and prompt augmentations.
First, we substituted the keyboard interaction with two physical buttons, to avoid accidental pressing of the touchpad or the other keys of the laptop.
Specifically, one button was registered as the ``talk'' button, allowing the user to ask questions to the LLM, while the other was used as the ``halt'' button to silence the system in case of long LLM answers.
Also, we have changed the speech recording interaction so that the users would need to keep the button pressed while recording their query.
We also enabled the possibility to ask questions even when the user is not pointing on the map.
In such case, only general questions on the map can be answered.

In the pointing interaction, we relaxed the constraints for the recognition of the pointing gesture to better detect when the user is pointing vertically, and we increased the ``gravity'' parameter of the map features.
We also added the automatic reading of accessibility information if the user keeps pointing to a street or an intersection. 
We enabled the possibility to point with both hands as well.
When two pointing hands are tracked, the system refers to the hand that is currently moving, ignoring the other one.
This approach allows a common interaction modality that BLV people use with TMs: to ``anchor'' a point on the map with one hand, and explore with the other, thus allowing to assess relative positions of map features with respect to the anchored point~\cite{zhao2021tactile}.

Concerning the navigation, since asking for the guidance instructions one at a time was still difficult for the participants, we introduced a real time navigation functionality called ``street-by-street navigation''.
This functionality automatically provides the next navigation instruction, once the user completes a navigation step.
If the user moves in the wrong direction during a step, the system notifies the user and recalculates the navigation from the new position.
The instructions also give distance indications in terms of the number of traversed blocks, and we modified the directions provided by the navigation to always refer to the cardinal directions of the map, making them fixed with respect to the user's reference frame (\textit{i.e.}, west would always be to the left of the user).
An important aspect of this function is that it is implemented locally by \system{} thanks to an additional tool call. This guarantees a more responsive interaction because the LLM initiates the navigation only but is not involved in the following navigation interactions.

Finally, we also introduced another guidance functionality, called ``fly-me-there guidance'', that guides the user to reach a point on the map without following the street graph.
This guidance modality provides verbal indications of the cardinal direction toward which the user should move the finger to minimize the distance to the target (a beaconing approach). 
This functionality was meant to be quicker for reaching a target position when it is not needed to know which path to take towards it.
This function was implemented with the same approach as the street-by-street navigation so it is run locally by \system{}.

\section{Prompt augmentations engineering}
\label{sec:engineering}


This section investigates LLM reliability in answering questions related to TMAPs (\textbf{Rq2}) and how the LLM answers can be improved by providing additional context information to each question (\textbf{Rq2.1}).
Before illustrating the process of prompt augmentations engineering, we explain the procedure adopted by our system to generate prompts (Subsection~\ref{subsec:engineering-context}). Then, we present the adopted methodology (Subsection~\ref{subsec:engineering-methodology}) and finally the results (Subsection~\ref{subsec:engineering-results}).

\subsection{Prompt augmentations}
\label{subsec:engineering-context}

Many LLMs currently available (including ChatGPT and Gemini) allow defining directives to guide the model's behavior through a textual description called \textit{system instructions} which is typically provided at the beginning of the conversation.
To support the embodied interaction proposed in this paper, we organize system instructions into 
\textit{answering instructions} and \textit{map contextual information}.
The answering instructions include a set of directives to specify the lexical context (e.g., ``I'm a blind person who needs help to navigate a new neighborhood.''), define the tool calls and set the tone for the conversation (e.g., ``Ensure that your answer is unbiased and does not rely on stereotypes.'').
The map contextual information, instead, consist of a description of the map, the area where it is located, possibly together with additional information, like a road graph, POIs, etc.

In addition to system instructions, our system also needs to augment each request with the ``prompt contextual data'' (PCD) that describe the context that is specific to a given user's prompt, such as the position indicated on the map at the time of the request.
PCD is generated from the contextual data through an algorithm called the ``PCD generator''.

\begin{figure}
    \centering
   \includegraphics[width=\textwidth]{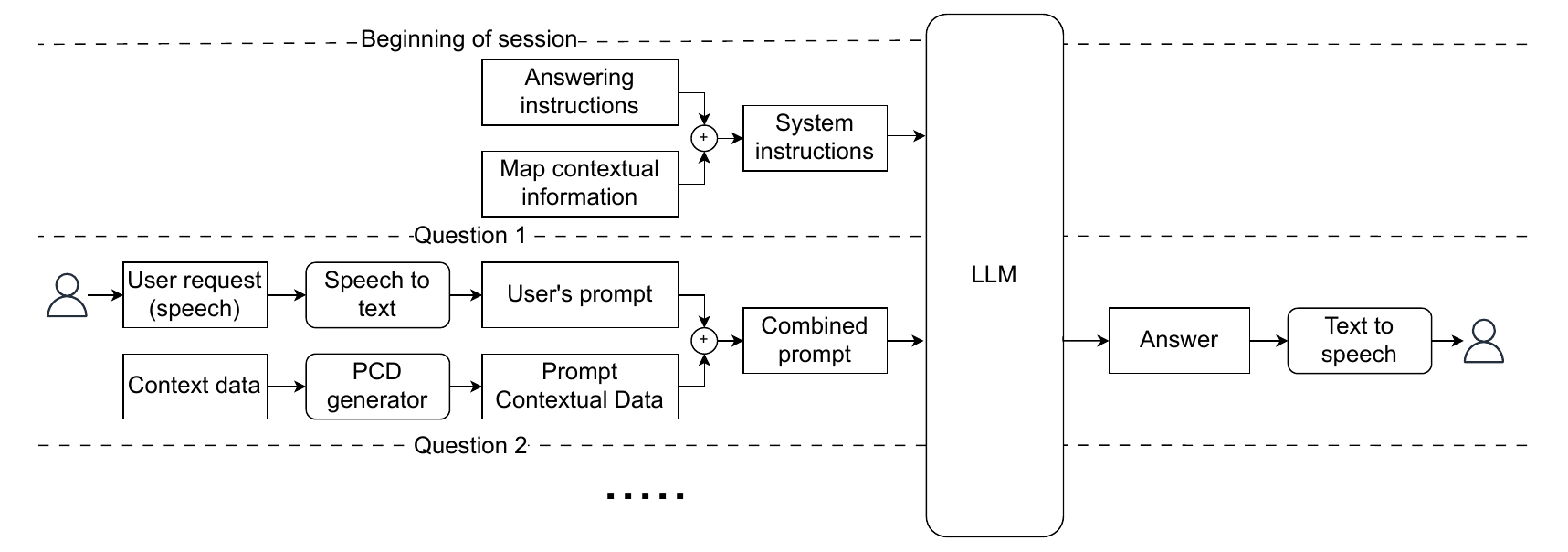}
    \caption{Data flow involving the LLM with focus on prompt augmentation.}
    \Description[Data flow of how user requests are processed to get an answer from the LLM.]{When the system is first started answering instructions and map contextual information are combined in system instructions are sent to the LLM. This does not produce any output. Then, everytime the user poses a verbal question, the audio recording is converted into text by the STT module and treated as the user's prompt. At the same time context data are computed and used to generate the PCD. The user's prompt and the PCD are combined in the combined prompt and sent to the LLM, which will produce an answer. The answer is played back to the user through TTS.}
    \label{fig:question_answer}
\end{figure}

Figure \ref{fig:question_answer} summarizes the data flow involving the LLM. \system{} communicates the system instructions to the LLM at the beginning of each session. Then, every time the user asks a verbal question, a speech-to-text module converts into a text (the user's prompt) and generates prompt contextual data from the current contextual data. The user's prompt and the prompt contextual data are concatenated to form the ``combined prompt'', which is then provided to the LLM. The LLM answer is read aloud to the user using a text-to-speech module.

\subsection{Methodology}
\label{subsec:engineering-methodology}

\begin{figure}
    \centering
   \includegraphics[width=0.9\textwidth]{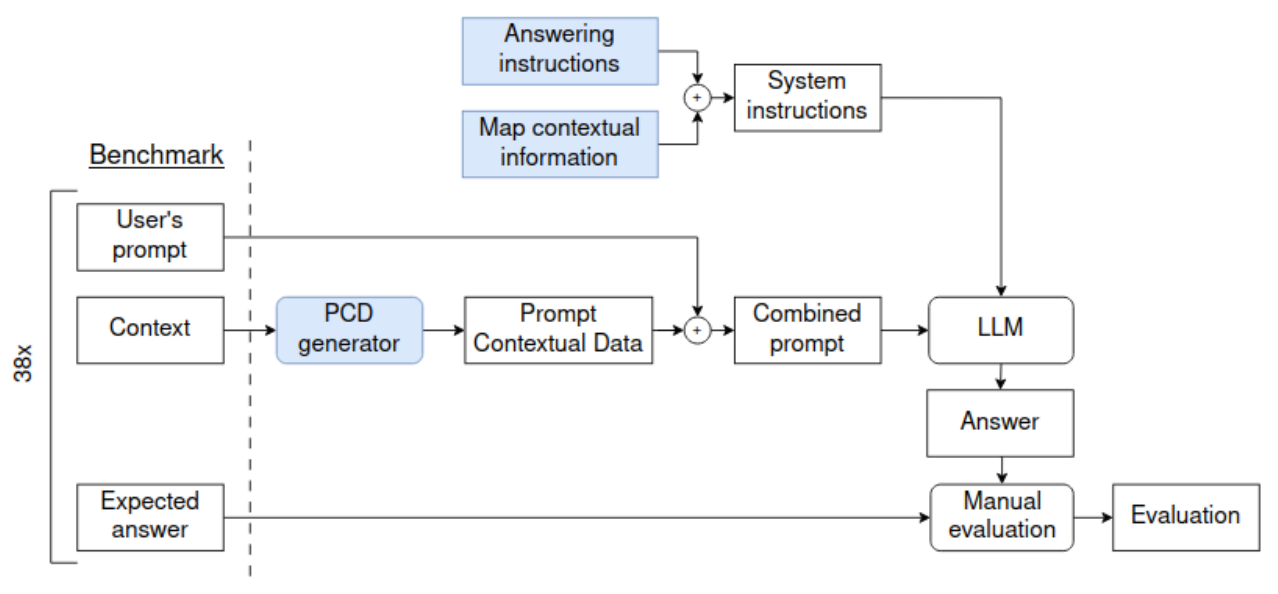}
    \caption{LLM query context engineering pipeline}
    \Description[Pipeline showing how benchmark queries are processed]{Each benchmark's query include a user's prompt, a context and an expected answer. The context is given to the PCD generator to generate prompt contextual data. These are combined with the user's prompt to form the combined prompt. At the same time answering instructions and map contextual information are combined in system instructions. System instructions and combined prompt are fed to the LLM to generate an answer, which is manually evaluated by comparing it with the query's expected answer. The result of the manual evaluation process is the evaluation of the query. This process is repeated once for each of the 38 queries of the benchmark and for each iteration.}
    \label{fig:benchmark_pipeline}
\end{figure}

The purpose of this design phase is to refine the answering instructions, the map contextual information and the PCD generator (\textbf{Rq2.1}) to improve the quality of the LLM's responses (\textbf{Rq2}). The methodology is illustrated in Figure~\ref{fig:benchmark_pipeline}.
The procedure is based on a benchmark of 38 user questions, each associated with a user prompt, a context, and the expected answer. An example of a request is: ``Tell me the roads parallel to this one'', the context is a position pointed on the map which correspond to a given road, and the answer contains the names of the two roads that are parallel to the one that is pointed.
The requests were designed based on the formative questionnaire (see Section~\ref{ssec:formativeQuestionnaire}) and follow the classification into landmark, route, and survey queries previously proposed in the literature \cite{brock2015interactivity}.
The complete list of questions is provided as Supplemental Material.

The process was iterative. In each iteration, we defined the answering instructions and the map contextual information, which jointly form the system instructions, and a PCD generator. We then provided to the LLM the system instructions and, for each request in the benchmark, the combined prompt, obtained by combining the user's prompt with the prompt contextual data generated from the context using the specified PCD generator.
Through a manual process, this response is evaluated against the expected response and classified into the following categories: Deceptively wrong, Not replying to question, Blatantly wrong, Partial or incomplete, Correct but not optimal, Correct.


This methodology makes it possible to compare different configurations of answering instructions, map contextual information and PCD generator measuring the results in objective terms.

\subsection{Results}
\label{subsec:engineering-results}

The iterative prompt-building process consisted of eight iterations. The first seven were conducted before the observational study described in Subsection \ref{ssec:firstPreliminaryStudy}, while the eighth took place before the final evaluation with users (Section~\ref{sec:evaluation}).
The benchmark was
designed for a map covering an area in New York City (this was used in the training phase of the evaluation, see Section~\ref{sec:evaluation}).
Table \ref{tab:benchmark_results} presents the results of each iteration. In the following, we discuss these results, grouped by structural similarities.

\begin{table}[ht]
\centering
\setlength{\tabcolsep}{0.2em}
\caption{Benchmark results. It: iteration, DW: deceptively wrong, NRW: not replying to question, BW: blatantly wrong, PI: partial or incomplete, CNO: correct but not optimal, C: correct}
\begin{tabular}{|c|l|l|l|cccccc|}
\hline
\multicolumn{1}{|l|}{\multirow{2}{*}{\textbf{It.}}} & \multirow{2}{*}{\textbf{\begin{tabular}[c]{@{}l@{}}Answering\\ instructions\end{tabular}}} & \multirow{2}{*}{\textbf{\begin{tabular}[c]{@{}l@{}}Map contextual\\ information\end{tabular}}} & \multirow{2}{*}{\textbf{\begin{tabular}[c]{@{}l@{}}Output of\\ PCD Generator\end{tabular}}} & \multicolumn{6}{c|}{\textbf{Results (\%)}} \\ \cline{5-10} 
\multicolumn{1}{|l|}{} &  &  &  & \multicolumn{1}{c|}{\textbf{DW}} & \multicolumn{1}{c|}{\textbf{NRW}} & \multicolumn{1}{c|}{\textbf{BW}} & \multicolumn{1}{c|}{\textbf{PI}} & \multicolumn{1}{c|}{\textbf{CNO}} & \textbf{C} \\ \hline
1 & \begin{tabular}[c]{@{}l@{}}Neighborhood  resident,\\ conversational,\\ detailed yet concise\end{tabular} & \begin{tabular}[c]{@{}l@{}}Textual description\\ of the map area\end{tabular} & \begin{tabular}[c]{@{}l@{}}Textual\\ description of\\ pointed position\end{tabular} & \multicolumn{1}{c|}{5.26} & \multicolumn{1}{c|}{28.95} & \multicolumn{1}{c|}{2.63} & \multicolumn{1}{c|}{0.00} & \multicolumn{1}{c|}{13.16} & 50.00 \\ \hline
2 & \multirow{4}{*}{\begin{tabular}[c]{@{}l@{}}As above +\\ rely only on the\\ provided info\end{tabular}} & \multirow{2}{*}{\begin{tabular}[c]{@{}l@{}}As above +\\ list of PoIs, map\\ coordinates and scale\end{tabular}} & \begin{tabular}[c]{@{}l@{}}Camera image\\ (map and hand)\end{tabular} & \multicolumn{1}{c|}{18.42} & \multicolumn{1}{c|}{0.00} & \multicolumn{1}{c|}{34.21} & \multicolumn{1}{c|}{0.00} & \multicolumn{1}{c|}{5.26} & 42.11 \\ \cline{1-1} \cline{4-10} 
3 &  &  & \begin{tabular}[c]{@{}l@{}}Synthetic image\\ (map+marker)\end{tabular} & \multicolumn{1}{c|}{21.05} & \multicolumn{1}{c|}{0.00} & \multicolumn{1}{c|}{26.32} & \multicolumn{1}{c|}{0.00} & \multicolumn{1}{c|}{15.79} & 36.84 \\ \cline{1-1} \cline{3-10} 
4 &  & \begin{tabular}[c]{@{}l@{}}As above + \\ local reference system+\\ road graph as JSON\end{tabular} & \multirow{3}{*}{\begin{tabular}[c]{@{}l@{}}Local coord.,\\ closest edge,\\ address,\\ distance\\ from nodes\end{tabular}} & \multicolumn{1}{c|}{10.53} & \multicolumn{1}{c|}{0.00} & \multicolumn{1}{c|}{10.53} & \multicolumn{1}{c|}{0.00} & \multicolumn{1}{c|}{15.79} & 63.16 \\ \cline{1-1} \cline{3-3} \cline{5-10} 
5 &  & \multirow{2}{*}{\begin{tabular}[c]{@{}l@{}}As in I4 but\\ road graph as text\end{tabular}} &  & \multicolumn{1}{c|}{2.63} & \multicolumn{1}{c|}{0.00} & \multicolumn{1}{c|}{2.63} & \multicolumn{1}{c|}{0.00} & \multicolumn{1}{c|}{13.16} & 81.58 \\ \cline{1-2} \cline{5-10} 
6 & \begin{tabular}[c]{@{}l@{}}As  above +\\ tool calls for\\ spatial reasoning\end{tabular} &  &  & \multicolumn{1}{c|}{0.00} & \multicolumn{1}{c|}{0.00} & \multicolumn{1}{c|}{2.63} & \multicolumn{1}{c|}{0.00} & \multicolumn{1}{c|}{7.89} & 89.47 \\ \hline
7 & \begin{tabular}[c]{@{}l@{}}As above +\\ street-by-street navigation\\ tool call for activating POI\end{tabular} & \multirow{2}{*}{\begin{tabular}[c]{@{}l@{}}As above + \\ specify if each node is\\ T and 4-ways\\ intersection\end{tabular}} & \multirow{2}{*}{\begin{tabular}[c]{@{}l@{}}As above +\\ type of closer\\ intersection\\ (T or 4 ways)\end{tabular}} & \multicolumn{1}{c|}{0.00} & \multicolumn{1}{c|}{0.00} & \multicolumn{1}{c|}{0.00} & \multicolumn{1}{c|}{0.00} & \multicolumn{1}{c|}{10.53} & 89.47 \\ \cline{1-2} \cline{5-10} 
8 & \begin{tabular}[c]{@{}l@{}}As above+\\ tool call for\\ navigation\end{tabular} &  &  & \multicolumn{1}{c|}{0.00} & \multicolumn{1}{c|}{0.00} & \multicolumn{1}{c|}{0.00} & \multicolumn{1}{c|}{0.00} & \multicolumn{1}{c|}{5.26} & 94.74 \\ \hline
\end{tabular}
\label{tab:benchmark_results}
\end{table}

\subsubsection{First iteration}


In the first iteration we relied solely on the pre-existing LLM knowledge to generate responses.
The answering instructions specified to the LLM that the user is blind and gave it the role of a long-time resident of the neighborhood. This was intended to encourage the LLM to provide details and create a friendly, conversational tone.
Additionally, answering instructions specified how to react to ambiguous or unclear queries - by asking for clarification - and to respond directly, with a detailed yet concise answer.
The map contextual information included only a brief textual description of the map, saying ``I am in New York, in the Empire State Building district''.
Meanwhile, the PCD generator was designed to compute a textual description of the user's position and current time, like, for example, ``at the intersection between Broadway and West 35th Street'' followed by ``The current time is: ...''.

The goal of this iteration was to explore the extent of the knowledge of the LLM. The benchmark revealed a $50.00\%$ rate of correct responses, with most correct answers concerning well-known POIs, such as the Empire State Building, or notable roads like Broadway. When the LLM encountered queries about less relevant POIs (e.g., Cooper Electrics, West 38th Street, New York City), it often acknowledged the lack of information ($28.95\%$ of responses). However, in some instances, it generated fabricated responses (hallucination): $5.26\%$ of responses were classified as deceptively wrong, and $2.63\%$ as blatantly wrong.
These findings highlighted the inadequacy of this approach (\textbf{Rq2}) and convinced us to explore different forms of map contextual data and policies for the PCD generator.

\subsubsection{Iterations 2 and 3}



Since the first iteration revealed the limitations of the LLM's knowledge, we aimed to enhance its performance in the second and third iterations by including in the map contextual data a list of POIs that includes shops and other public places extracted from a public source\footnote{\url{apidocs.geoapify.com}}.
This list was presented to the LLM in textual form, reporting, for each POI its location (in terms of latitude and longitude) and descriptive information, like name, category, extended description, opening hours, accessibility, etc.
We also provide a reference system: the four map corners and the scale (``The scale of the map is: $60$m: $1$cm'').
The answering instructions were also expanded to prevent the LLM from accessing its pre-existing knowledge by saying ``Stick to the provided information: when information is insufficient to answer a question, respond by acknowledging the lack of an answer and suggest a way for me to find one.''

The PCD generator varied across the two iterations. In the second iteration, it produced the RGB image acquired from the camera. This image frames the map and the user hand that is aiming the target of the map.
Instead, in the third iteration, the PCD generator processes the camera image by using a combination of algorithms based on machine learning and rules, and computes the pointed position. Then, it augments a pre-existing image of the map with a marker representing the user's position and returns the result. 

We expected these changes to improve the performance, instead benchmark results showed only $41.11\%$ of correct answer for the second iteration and $36.84\%$ for the third, less than in the first iteration. Similarly, the percentage of responses classified as deceptively wrong and blatantly wrong increased, rising to $18.42\%$ and $34.21\%$ for the second iteration, and $21.05\%$ and $26.32\%$ for the third.
Our interpretation of these results is that the LLM is unable to perform spatial reasoning based on the provided images. This is also reported in the OpenAI documentation \cite{openaiLimitations}.

\subsubsection{Iterations 4 and 5}

In order to address the issues that arose in iterations 2 and 3, 
we extended the map contextual information to explicitly represent the map's road network as a graph, where nodes indicates intersections between streets and edges represent street segments.
This representation allows to include information about the accessibility of nodes and edges, such as the presence of crosswalk signals at intersections or ongoing roadwork on specific street segments. Additional contextual data about the map, including its name and nearby locations (north, east, south, and west), are provided to the LLM.
We also introduced a local Euclidean reference system. Nodes are placed within this system in such a way that the distance between any two nodes matches the real-world distance, in meters, between the corresponding intersections.
A POI's position is represented in various formats: in terms of the local reference system, the address, and the closest edge on the graph. The intuition is that, while in principle the latter two formats can be derived from the local position, the conversion can be subject to errors. With the proposed solution the LLM can select the format that is more practical for a given task, hence improving the quality of the answer.
In the fourth iteration, the graph was provided to the LLM in a JSON format along with a final brief explanation of its structure. In the fifth iteration, it was provided as text, where graph elements were alternated with their descriptions to create a more cohesive description.

Similarly to a POI's position, the pointing position is also reported in various formats in addition to the local reference system, including the closest edge, the edge name and the distance from the edge nodes as in ``[...] the closest point on the road network is on edge n1 - n2, which is part of West 38th Street, between 5th Avenue and 6th Avenue. I'm at a distance of 100 m from the intersection with 6th Avenue and 210 m from the the intersection with 5th Avenue.''


Benchmark results showed the best results so far. Both iterations achieved more correct answers than the previous ones: $63.16\%$ and $81.58\%$ respectively.
The text based representation (iteration 5) yielded particularly good performance, with only one answer evaluated as deceptively wrong and one as blatantly wrong.
However, the LLM continued to struggle with some questions that requires spatial reasoning, like, for example, ``How can I come here from Solle Spa?'' and ``How far is Solle Spa on foot?''. In the former, the LLM completed the answer with wrong accessibility information, while in the latter, it slightly miscalculated the distance.




\subsubsection{Iterations 6 and 7}
\label{sssec:iterations6and7}

As a solution to the LLM poor spatial reasoning abilities, we decided to leverage OpenAI's GPT-4o's support for tool calls to complete specific tasks with ad-hoc procedures.
Starting from iteration six, we introduced a set of tool calls to compute the following spatial functions: the distance between two points (or between a point and a POI), whether a given position is close to a given POI, a list of POIs close to a given point, a list of instructions to navigate between two points (or between a point and a POI).
Thanks to the use of these tool calls, 
we increased the number of correct responses to $89.47\%$, with only a single response classified as blatantly wrong and no deceptively wrong answers.


Iteration seven introduces the distinction between T and 4-way intersections both in the map contextual information and in the output of the PCD generator.
Answering instructions were improved as well,
by instructing the LLM to provide the first step only in the navigation, and to invite the user to ask for the next step in a follow-up request. Also, we added a tool call make a POI discoverable on the map when pointed to.
All these changes led to further improvements in benchmark results, with all responses classified as either correct ($89.47\%$) or correct but not optimal ($10.53\%$). Non-optimal responses only were primarily due to verbose answers to questions related to navigation. The problem is that, when providing navigation instructions, the LLM provides all the instructions at the same time, in a list that is difficult to remember.




\subsubsection{Iteration 8}


As reported in Section~\ref{ssec:secondPreliminaryStudy}, after the preliminary tests, we implemented two new navigation modalities.
To support them, we added two new tool calls, hence adding their definition to the answering instructions as well as the text required to explain to the LLM when to activate one or the other. 
For instance, if the user asks ``guide me to ...'', fly-me-there guidance should be activated, while, if they ask ``navigate me to ...'' or ``give me directions to ...'', street-by-street navigation should be activated. When the context of the question is unclear, the LLM should directly ask the user which option they prefer.
This expected behavior is explained to the LLM through example questions paired with their expected answers, which are included in the answering instructions to encourage standardized responses to the most frequent queries.
%
%
In the eighth iteration, the benchmark results achieved a score of $94.74\%$ correct responses and $5.26\%$ correct but not optimal on the New York City map.

\section{Experimental Methodology}
\label{sec:evaluation}
To assess the proposed interaction paradigm and the system (\textbf{Rq4}), we conducted a user study with ten BLV participants.
The study was approved by the ethics committee of \anonymized{The Smith-Kettlewell Eye Research Institute, with the approval number: XXXXXX}\footnote{Planning, conduct, and reporting of the work are consistent with the general principles enunciated in~\cite{acmetics}.}.
This section reports the experimental methodology of the study, while Section~\ref{sec:results} reports the results.

\subsection{Apparatus, Setting, and Stimuli}
\label{subsec:evaluation-apparatus}
The study was conducted in a laboratory at \anonymized{The Smith-Kettlewell Eye Research Institute}, with the participant sitting at a table in a quiet room.
An Apple MacBook Pro M2 laptop running \system{} was used for the experiments.
An external camera was connected to the laptop and mounted above the tactile map to track the participant's hand gestures and the pointed location.

Additionally, two Bluetooth buttons\footnote{AbleNet's Blue2 FT - \url{https://www.ablenetinc.com/blue2-ft}} were connected to the laptop for input.
One button, identified by a small bump dot, served as the ``talk'' button, while the other one was used as the ``halt'' button.
The buttons could be freely placed by the participant according to their preference.
The participant's voice was recorded via the laptop's built-in microphone, while system messages and responses from the LLM were played through the laptop's built-in speakers.

\begin{figure}
    \centering
   \includegraphics[width=0.4\textwidth]{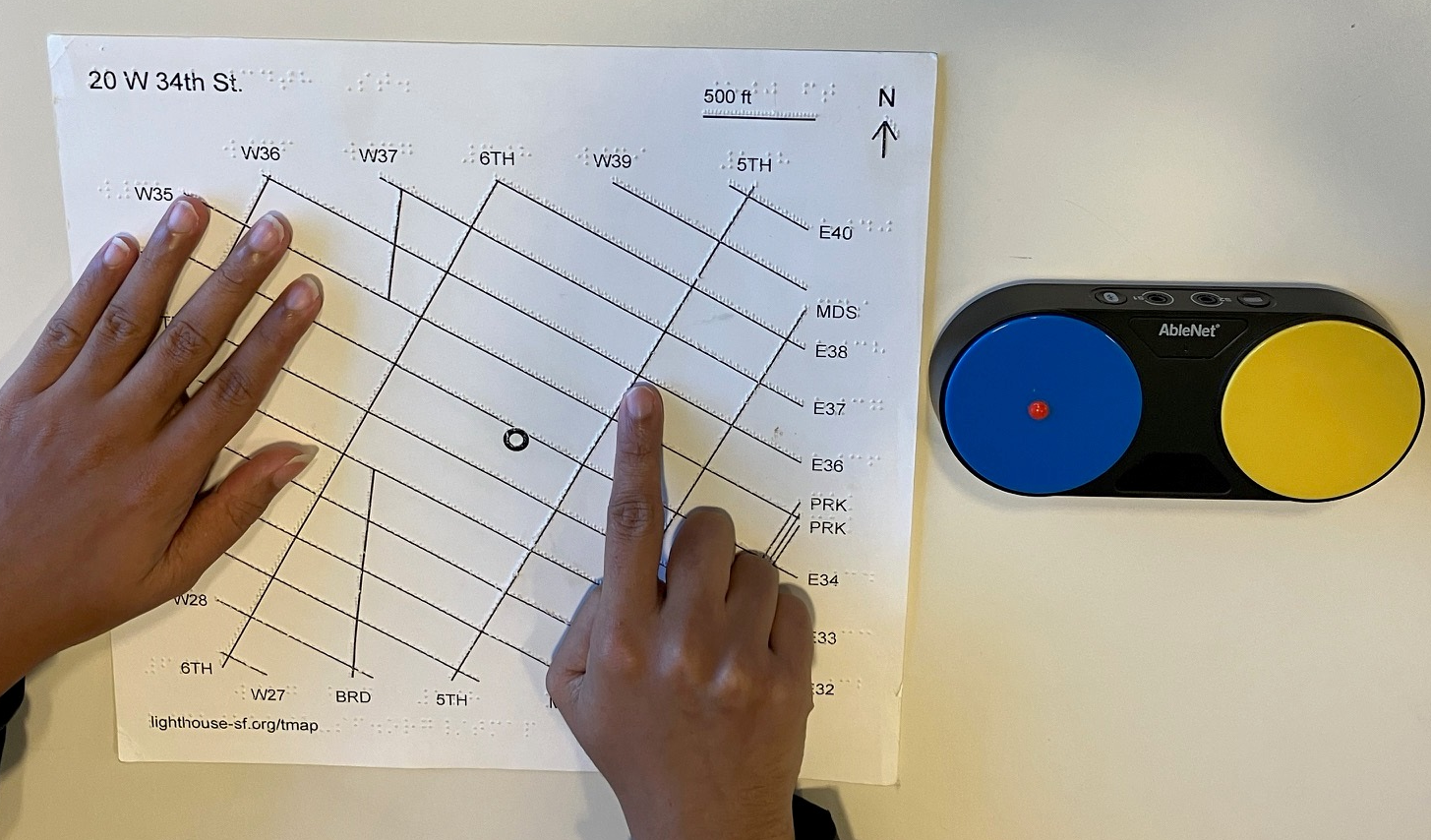}
    \caption{Participant shown interacting with \system{}, with Bluetooth buttons nearby.}
    \label{fig:setup}
    \Description[Participant's hands interact with a map, with two Bluetooth buttons nearby.]{A tactile map of Detroit is placed on a white table. The hands of a participant are shown while interacting with \system{}: the left hand is open and placed on the map; the right hand is pointing at an intersection on the map.
    Two Bluetooth buttons are placed on the right of the map: the one on the left is the ``talk'' button, it is blue and has a small bump dot on it; the other one is the ``halt'' button and it is yellow.}
\end{figure}

As stimuli, we used two TMs: one of the Empire State Building neighborhood in Manhattan, New York City, NY, USA, and another of the Conant Gardens neighborhood in Detroit, Michigan, USA.
The New York map was used for training purposes, while the Detroit map was used for the actual tasks.
The maps were purchased from the LightHouse for the Blind and Visually Impaired\footnote{\url{https://lighthouse-sf.org/tactile-images-maps/tmap}}.

\subsection{Experimental Design and Metrics}
\label{ssec:experimentalDesign}
The evaluation is conducted with a mixed-method approach that combines an observational study, a usability evaluation and a final interview.
For the observational study, we collected video and audio recordings from the camera framing the map and the participants' hands.
We collected additional system logs and debug information, including conversational interaction transcriptions, the position on the map pointed by the participant, and information regarding the computer vision functions (\textit{e.g.}, whether the hands are in the pointing position).
The final interview included a series of open questions regarding feedback on the pointing interaction, interaction with the LLM, the usefulness of the two navigation modes, and general reflections on the system's strengths and weaknesses.
The complete list of questions is reported in Supplemental Material.

The visual study and the interview are analyzed through reflexive thematic analysis~\cite{braun2019reflecting}.
Both inductive and deductive approaches were used.
The inductive approach was based on the observational analysis of video and audio recordings of the experiments.
Three researchers jointly viewed and coded the recordings of one participant, registering emotional responses, behavioral and interaction patterns, and in particular disconnects between their natural way of interacting through touch and conversation, with respect to the designed interaction paradigm.
Then the rest of the videos were randomly split between the three researchers, with each video coded by one primary coder, and codes integrated with additional observations from a secondary coder.
During a follow up meeting, codes were reviewed and adjusted.
Uncertainty or disagreements among the coders were addressed through reviewing and discussion of the relevant parts of the recordings, until consensus was reached.
The deductive approach derived theory-driven themes and codes, starting from the examined research questions (Section~\ref{sec:intro}), the system and interaction design (Section~\ref{sec:design}), and the topics investigated in our final questionnaire.
These topics were tightly related to the system functionalities and the proposed interaction modality.
The codes were finally merged and reviewed by the researchers, grouping them into subthemes and themes.

The system usability was assessed using the the System Usability Scale (SUS) questionnaire \cite{brooke1996sus}.
The SUS questionnaire follows the adaptation proposed by Brock et al. \cite{brock2015interactivity}, which modifies the seventh question to read, ``I would imagine that most BLV people would learn to use this system very quickly''.
This change prompts participants to evaluate the system with its intended audience in mind.
We descriptively analyzed the overall score and single items, comparing them to benchmarks available in the literature~\cite{bangor2009determining,lewis2018item}.




\subsection{Experimental Protocol}
\label{subsec:evaluation-protocol}

The study was organized into five main phases.
%
In the first phase we gathered participants' demographic information, along with data on their visual condition, expertise and habits with braille, TMs, LLMs and digitally augmented tactile supports.
%
In the second phase, we provided a brief introduction to the study, outlined the system's functionalities, and encouraged participants to ask questions at any point and to adopt a think-aloud approach throughout the study
(although we also suggested participants not talk to the supervisor while asking verbal questions to the LLM).

The third phase was the training, during which we gradually introduced participants to the system using the New York map. We first explained how to point to a location on the map to receive feedback on their position and asked them to complete basic tasks, such as identifying a road and obtaining additional details about it.
Next, we demonstrated how to interact with the LLM by asking questions and prompted participants to ask questions themselves, such as inquiring about restaurants on the map.
Finally, we explained how to request directions to specific locations, practicing both fly-me-there guidance and street-by-street navigation.
During this phase, the supervisor assisted participants whenever they encountered difficulties completing a task.

The fourth phase was the formal evaluation, during which each participant was asked to complete ten tasks using the TMAP of a neighborhood in Detroit:
\begin{enumerate}[label=\arabic*.]
\item Explore the map freely for one minute, and use the question and answer mode to ask something. [max 2 minutes]
\item You will be staying at the Sheraton Commander Grand Lake hotel. Is this hotel on the map? [max 2 minutes]
\item Is there free Wi-Fi at the hotel? [max 2 minutes]
\item Since the hotel has a long name, ask the system to remember it as ``my hotel''. [max 2 minutes]
\item You want to find a nice restaurant where you can meet your friends for a late evening dinner. Check if there is an Italian restaurant. [max 2 minutes]
\item Check if the restaurant is open at 10 PM. [max 2 minutes]
\item Check how long it would take you to walk there from the hotel. [max 2 minutes]
\item You decide to go by taxi. Ask the system to help you point to the restaurant on the map. Once your finger is pointing there, please keep it pointing there for the next two tasks. [max 3 minutes]
\item You want to check what entertainment options are near the restaurant. [max 2 minutes]
\item After the dinner you want to go back to the hotel. Ask the system to navigate you there. [max 5 minutes]
\end{enumerate}
Each task had a set time limit, after which the supervisor would proceed to the next task.
During this phase, the supervisor was not permitted to assist participants in task completion.
However, if a task depended on the successful completion of a previous one, the supervisor showed the participant how to complete it. 
The last phase of the study collected SUS answers and additional subjective feedback and comments from the participants regarding the whole study.
Additional details about the protocol are reported in Supplemental Material.

\subsection{Participants}
\label{subsec:evaluation-participants}

The study involved ten participants, whose demographic data are reported in Table~\ref{tab:evaluation-demographic}.
The inclusion criterion for this study required the participants to be blind or legally blind.
The exclusion criteria were: having familiarity with the area represented in the Detroit map, having participated in the focus group or the preliminary studies (see Section~\ref{sec:design}), or the presence of other impairments that could influence the study.

Seven participants self-identified as female, while the remaining participants identified as male. Participants' ages ranged from $30$ to $78$ ($56.11 \pm 15.73$)\footnote{We use the notation $x \pm y$ to denote average $x$ and standard deviation $y$.}.
We also collected data on participants' self-reported expertise with braille, TMs, and LLMs on a Likert-like scale (``None'', ``Low'', ``Mid'', ``High").
Notably, \textbf{P12}, \textbf{P15}, and \textbf{P20} also indicated they had prior experience in training other BLV individuals.

Finally, we inquired about participants' past experiences with digitally augmented tactile materials. 
Participants were allowed to provide open-ended responses.
\textbf{P12}, \textbf{P14}, \textbf{P16}, and \textbf{P20} had little prior experience with digitally augmented tactile materials, primarily during previous experiments at \anonymized{The Smith-Kettlewell Eye Research Institute}.

\begin{table}[ht]
\centering
\setlength{\tabcolsep}{0.2em}
\caption{Participants in the main study. LP: light perception, FP: form perception, DATS: digitally augmented tactile supports}
\begin{tabular}{|c|c|c|c|c|c|c|c|c|c|c|}
\hline
\multirow{2}{*}{\textbf{PID}} & \multirow{2}{*}{\textbf{Age}} & \multirow{2}{*}{\textbf{Gender}} & \multicolumn{3}{c|}{\textbf{Visual Impairment}} & \multirow{2}{*}{\textbf{TBLV}} & \multicolumn{3}{c|}{\textbf{Expertise}} & \multirow{2}{*}{\makecell{\textbf{Experience}\\ \textbf{with DATS}}} \\
\cline{4-6}
\cline{8-10}
& & & \textbf{Description} & \textbf{Onset} & \textbf{Residual} & & \textbf{Braille} & \textbf{TMs} & \textbf{LLMs} & \\ \hline
\textbf{P11} & 39 & F & ROP~\cite{hellstrom2013retinopathy} & birth & 20/200 & No & High & Low & Low & No \\ \hline
\textbf{P12} & 68 & M & Glaucoma & birth & Yes & Yes & Mid & Low & Mid & Yes \\ \hline
\textbf{P13} & 52 & M & RP~\cite{hartong2006retinitis} & age 7 & LP\&FP & No & None & Mid & High & Yes \\ \hline
\textbf{P14} & 63 & F & Blindness & age 1 & No & No & High & Mid & None & Yes \\ \hline
\textbf{P15} & 78 & M & Blindness & birth & No & Yes & High & Mid & None & Yes \\ \hline
\textbf{P16} & 43 & F & Uveitis~\cite{durrani2004uveitis} & age 5 & No & No & None & Mid & High & Yes \\ \hline
\textbf{P17} & N/A & F & Genetic mutation & birth & LP & No & Low & Mid & None & No \\ \hline
\textbf{P18} & 30 & F & ROP & birth & LP\&FP & No & High & Low & None & No \\ \hline
\textbf{P19} & 56 & F & LCA~\cite{den2008leber} & birth & LP & No & High & High & High & Yes \\ \hline
\textbf{P20} & 76 & F & ROP & birth & LP\&FP & Yes & High & Mid & None & Yes \\ \hline
\end{tabular}
\label{tab:evaluation-demographic}
\end{table}

\section{Experimental Results}
\label{sec:results}
All participants completed the experiment as expected, with an experiment duration of about one hour.
All participants were able to complete all the tasks in the formal evaluation, with the following exceptions: in task 8, \textbf{P12} erroneously searched for the hotel (instead of the restaurant).
Also, \textbf{P15} and \textbf{P16} were not able to instruct the LLM to save the bookmarks.
\textbf{P17} requested not to record the audio; the results of the final questionnaire and SUS with \textbf{P15} were lost due to technical problems.

\subsection{Thematic analysis}
The thematic analysis identified $4$ main themes through the deductive process, related to the \textit{System} and its functionalities (\textit{Pointing}, \textit{Conversation} and \textit{Navigation}).
The inductive process, instead, identified $5$ themes related to: subjective \textit{Perception} of system aspects or functionalities, their \textit{Use Cases}, the \textit{Naturalness} of the interactions, \textit{Desiderata} regarding existing or new functionalities, and how users address \textit{Fault and Recovery}.

We reflected on the fact that the two categorizations, stemming respectively from the inductive and the deductive approaches, are mutually orthogonal. 
Discussing on how to present the themes, we have realized that raising the deductive themes to main themes and for each having the inductive themes as sub-themes supports a clearer presentation of the findings.
Thus, our presentation follows this rationale.

\subsubsection{\textit{System}}

\paragraph{Perception}
Many participants expressed appreciation for \system{} (\textbf{P12}, \textbf{P13}, \textbf{P14}, \textbf{P16}, \textbf{P18}, \textbf{P19}, \textbf{P20}).
In particular, \textbf{P14} was impressed by the capabilities of the system and \textbf{P20} reported:
\begin{quote}
    \textit{``Would have liked to have done more! Want to use the system more!''}
\end{quote}

\paragraph{Use Cases}
Some criticisms were reported as well: \textbf{P12}, \textbf{P13} and \textbf{P16} commented that the system would not be usable on the go while \textbf{P19} and \textbf{P20} raised the problem of map creation, both in terms of cost and dependence on others to make the maps.
Despite this, all participants reported at least one use case in which they would use \system{}, including exploration (\textbf{P11}, \textbf{P12}, \textbf{P17}, \textbf{P18}), navigation (\textbf{P13}, \textbf{P14}, \textbf{P16}, \textbf{P19}, \textbf{P20}) and retrieval of accessibility features (\textbf{P11}). For instance \textbf{P12} would use \system{} to preview an unfamiliar environment.

\paragraph{Naturalness}
All participants explored maps with an open hands, some also with two (\textbf{P11}, \textbf{P13}, \textbf{P15}, \textbf{P17}, \textbf{P18}, \textbf{P19}, \textbf{P20}).
In particular, \textbf{P17} explored systematically starting from the map perimeter and from the center.
This interaction appears to be natural and it is one of the advantages of the system with respect to approaches using touch surfaces that react to all touches.
However, \textbf{P20} notes that it does take time to learn to use the system.

\paragraph{Desiderata}
System's ability to be used in diverse contexts, in particular in mobility, was requested by some (\textbf{P13}, \textbf{P14}, \textbf{P17}), while others suggested integration with more portable form factors such as smart glasses (\textbf{P16}) or link with other navigation tools such as BlindSquare (\textbf{P12}).
\textbf{P19} foresees possible generalization to other types of tactile materials (\textit{e.g.}, diagrams in education).
Supporting personalization based on user preferences and interests was also suggested by \textbf{P12} (\textit{e.g.}, to tag favorite places) and by \textbf{P16} (\textit{e.g.}, to change distance units).

\paragraph{Fault and Recovery}
Despite the appreciation for the system, trust in provided information is one concern that BLV people have when interacting with technology~\cite{muller2022traveling}.
Indeed, \textbf{P19} raised the issue of wanting to cross-check the information provided by the system (\textit{e.g.}, using Google Maps).

\subsubsection{\textit{Pointing}}

\paragraph{Perception}
All participants appreciated the pointing functionality, and most praised its accuracy (\textbf{P11}, \textbf{P13}, \textbf{P14}, \textbf{P16}, \textbf{P17}, \textbf{P19}).
Similarly, all highlighted the clarity of the provided information.
In particular the first use of the functionality triggered a strong emotional response in \textbf{P11}.

\paragraph{Use Cases}
All participants reported the usefulness of the information provided, which was regarded to be complete by most (\textbf{P11}, \textbf{P14}, \textbf{P16}, \textbf{P18}, \textbf{P19}).
However, \textbf{P13} highlighted that real world use is needed to assess this aspect,
while \textbf{P12}, \textbf{P16}, and \textbf{P20} were similarly impressed by the detail of the provided feedback. Specifically, \textbf{P12} commented:
\begin{quote}
    ``\textit{They are giving terrain information as well! That's interesting.}''
\end{quote}
In contrast, some found the information redundant (\textbf{P11}, \textbf{P18}) or sometimes redundant (\textbf{P12}, \textbf{P19}). In particular \textit{P18} felt overwhelmed by the all the provided information.

\paragraph{Naturalness}
During training, the pointing gesture took some time to learn by many participants (\textbf{P11}, \textbf{P12}, \textbf{P16}, \textbf{P17}, \textbf{P20}), or was not perceived as natural (\textbf{P13}, \textbf{P15}, \textbf{P18}).
However, most had no problems with it once learned (as highlighted by \textbf{P12}),
with the exception of \textbf{P20} who had some problems with it also during test.
Once learned, participants would naturally use pointing gesture to explore alternatively to the open hand exploration (\textbf{P12}) or in combination with it (\textbf{P15}, \textbf{P17}).
Concerning the two-hand pointing, \textbf{P13} expressed appreciation for it.
However, most participants did not use this functionality in practice (\textbf{P11}, \textbf{P12}, \textbf{P13}, \textbf{P14}).
Indeed \textbf{P18} found it confusing, while \textbf{P14}, \textbf{P15}, \textbf{P20} had difficulties when they tried using it.
However, \textbf{P20} did eventually use this functionality spontaneously.

\paragraph{Desiderata}
In addition to using the pointing functionality for accessing street information, some participants (\textbf{P11}, \textbf{P13}, \textbf{P14}, \textbf{P15}, \textbf{P16}, \textbf{P19}) pointed to braille labels, expecting to receive their description.
This suggests that the system should also be able to describe braille labels.
Despite the appreciation for pointing feedback, in some occasions, participants were annoyed by it (\textit{e.g.}, when talking to the experimenter), hinting that pausing audio feedback (\textbf{P12}, \textbf{P13}, \textbf{P16}, \textbf{P18}, \textbf{P20}) should be a functionality.
Some also wanted to turn off the ``crickets'' sound or to personalize it (\textbf{P11}, \textbf{P16}).

\paragraph{Fault and Recovery}
The ``Crickets'' sound was nonetheless found to be a welcome indicator of the functioning of the system (\textbf{P16}):
\begin{quote}
    ``\textit{At least crickets tell you it's still on. If an app is quiet, I don't know what to do. SeeingAI gives feedback but sometimes it's silent for a while, now what do I do?}''
\end{quote}
Similarly the repetitions of audio information were appreciated when information is missed the first time (\textbf{P17}).
The pointing itself was a potential source of faults as the participants had no clue regarding the area framed by the camera which, if wrongly positioned, could cause issues that participants could not address autonomously (\textbf{P12}, \textbf{P15}).
Occlusion caused by participants' own body (e.g., one hand in front of the other) was also a potential source of pointing recognition errors, which were hard for the participants to understand (\textbf{P11}).

\subsubsection{\textit{Conversation}}

\paragraph{Perception}
Many participants appreciated the conversation functionality and its ability to understand their questions (\textbf{P11}, \textbf{P12}, \textbf{P13}, \textbf{P14}, \textbf{P16}, \textbf{P17}, \textbf{P20}).
In particular \textbf{P11} expressed enthusiasm when first interacting through conversation.
In general, all participants considered the provided answers clear,
complete,
correct,
concise,
and useful.
Concise responses are also the reason why all participants except \textbf{P15} and \textbf{P19} never used the halt button to interrupt a response. 
However, the system could take time to respond to user questions, which could cause annoyance, as expressed by \textbf{P19}, or could result overwhelming if too much information is provided (\textbf{P12}).

\paragraph{Use Cases}
Participants commented on some use cases for which the conversational functionality was considered useful.
These include knowing restaurants (\textbf{P14}), entertainment options (\textbf{P13}) and other POIs (\textbf{P19}) near the pointed position.
Even just knowing that a POI is on the map is useful (\textbf{P12}), 
and getting the distance and walking time between two POIs is appreciated as well (\textbf{P13}, \textbf{P17}).
Finally, the ability to remember POIs as bookmarks was also appreciated (\textbf{P11}, \textbf{P16}, \textbf{P17}, \textbf{P18}).
In particular, \textbf{P17} said:
\begin{quote}
    ``\textit{I don't have to find the address again and again. Really loved that!}''
\end{quote}

\paragraph{Naturalness}
For some participants, ``push-to-talk'' interaction was not natural.
Indeed, some started talking without pushing the button (\textbf{P11}, \textbf{P13}, \textbf{P14}, \textbf{P15}, \textbf{P16}, \textbf{P19}).
Others pressed the button but did not keep it pushed (\textbf{P12}, \textbf{P13}, \textbf{P15}, \textbf{P16}, \textbf{P19}, \textbf{P20}).
Others again took some time to learn coordinating the pointing hand and the hand pushing the button (\textbf{P14}).
Also, the physical buttons we used for the experiment was too sensitive and most participants had problems to keep it pressed.
This challenge was unexpected and required some training (\textbf{P13}, \textbf{P14}, \textbf{P15}, \textbf{P16}, \textbf{P17}, \textbf{P19}, \textbf{P20}).
After some practice in the use of the ``push-to-talk'' button, most participants interacted through the conversational interface without issues.
In particular \textbf{P12} noted that it is easy to be understood by the system despite phrasing errors (see \textit{Fault and Recovery}).
However, some participants forgot how to ask for direction (\textbf{P15}) or how to save a bookmark (\textbf{P13}).
Also, initially \textbf{P15} asked incomplete questions (\textit{e.g.}, sentences without verb).
Similarly, \textbf{P19} had to think carefully about the phrasing and \textbf{P18} commented:
\begin{quote}
    \textit{``It took a bit of practice to figure out how to make myself sound clear to it''}
\end{quote}

\paragraph{Desiderata}
Participants requested various additional information to be made available through conversational interaction.
Some requested details about accessibility (\textbf{P13}, \textbf{P17}, \textbf{P19}), others about POIs facilities (\textbf{P14}, \textbf{P15}, \textbf{P20}) and others about public transportation (\textbf{P12}, \textbf{P13}, \textbf{P17}).

\paragraph{Fault and Recovery}
In most of the cases, if the participant's request was not correctly understood by the conversational interface, the problem was solved by reformulating the question (\textbf{P11}, \textbf{P13}, \textbf{P16}, \textbf{P18}, \textbf{P19}).
Participants appreciated when LLM recovered from minor pronunciation or formulation mistakes (\textbf{P11}, \textbf{P12}, \textbf{P14}, \textbf{P15}, \textbf{P19}, \textbf{P20}).
For example, \textbf{P15} was surprised that the LLM could correctly understand a prompt that he had the feeling was not clearly formulated:
\begin{quote}
    ``\textit{Very hip. Especially since I mess with your wording. It doesn't seem to care. It gets the essence of what I'm driving at.}''
\end{quote}
One exception was with the requests to remember the bookmarks which, if not phrased correctly, were interpreted by the LLM as attempts to store personal information and therefore were not processed (\textbf{P15}, \textbf{P19}).
We highlight that, since \system{} uses LLM for conversation support, hallucinations are also possible.
This however is mitigated by our prompt augmentations methodology (see Section~\ref{sec:engineering}).
Indeed we observed a single case of hallucination, when the system did not report nearby POIs (\textbf{P15}).  

\subsubsection{\textit{Navigation}}

\paragraph{Perception}
Some participants were thrilled when first using navigation functionalities (\textbf{P11} and \textbf{P20} for fly-me-there, \textbf{P13}, \textbf{P14}, \textbf{P20} for street-by-street).
However, these functionalities resulted hard to use, causing frustration (\textbf{P11} for fly-me-there, \textbf{P12}, \textbf{P14}, \textbf{P15}, \textbf{P20} for street-by-street).
Despite this, most participants managed to complete the navigation tasks, after some learning, as highlighted by \textbf{P15}.

\paragraph{Use Cases}
Specific use cases for the two navigation modalities were clear to the participants: street-by-street is preferable for learning routes (\textbf{P13}, \textbf{P14}, \textbf{P16}, \textbf{P17}, \textbf{P18}, \textbf{P20}), while fly-me-there gives a general direction of a POI (\textbf{P12}, \textbf{P13}, \textbf{P16}) or when getting to a location by car or transportation (\textbf{P17}).

\paragraph{Naturalness}
While street-by-street was generally more useful, it was also more difficult (\textbf{P11}, \textbf{P14}, \textbf{P18}, \textbf{P20}).
The differences between the two interactions sometimes confused the participants that followed the streets in fly-me-there (\textbf{P11}, \textbf{P12}, \textbf{P14}, \textbf{P16}, \textbf{P17},\textbf{P18}, \textbf{P19}, \textbf{P20}), or went over them in street-by-street navigation.
Another interaction problem was due to the fact that some participants were not proficient with cardinal directions, in particular the diagonal ones (\textit{e.g.}, "North-east'') that were common in the New York City map (\textbf{P11}, \textbf{P12}, \textbf{P13}, \textbf{P14}, \textbf{P16}, \textbf{P19}, \textbf{P20}).
%
In several cases the participants moved beyond the destination point during navigation due to a system delay in updating the position (\textbf{P11}, \textbf{P13}, \textbf{P14}, \textbf{P16}, \textbf{P17} \textbf{P18}, \textbf{P19}).
This also resulted in too many messages being provided (\textbf{P15}).

\paragraph{Desiderata}
The system delay (\textbf{P13}, \textbf{P16}, \textbf{P19}) and generally the reliability during navigation (\textbf{P14}, \textbf{P20}) were also the main requests for improvement for this functionality.
Furthermore, to address the lack of proficiency with cardinal directions, \textbf{P11} proposes to use clockface directions.

\paragraph{Fault and Recovery}
System delays would cause some participants to start moving erratically due to frustration (\textbf{P13}, \textbf{P16}, \textbf{P17}, \textbf{P20}).
This situation was often addressed by slowing down (\textbf{P11}, \textbf{P14}, \textbf{P16}, \textbf{P17}) or by backtracking (\textbf{P16}).
In sporadic cases, during navigation, there were problems in computing the homography due to the fact that users were covering a large part of the map with their hands. This in turn resulted in position computation errors, ultimately preventing the user from finding the destination (\textbf{P19}).

\subsection{Usability evaluation}
\begin{figure}
    \centering
   \includegraphics[width=\textwidth]{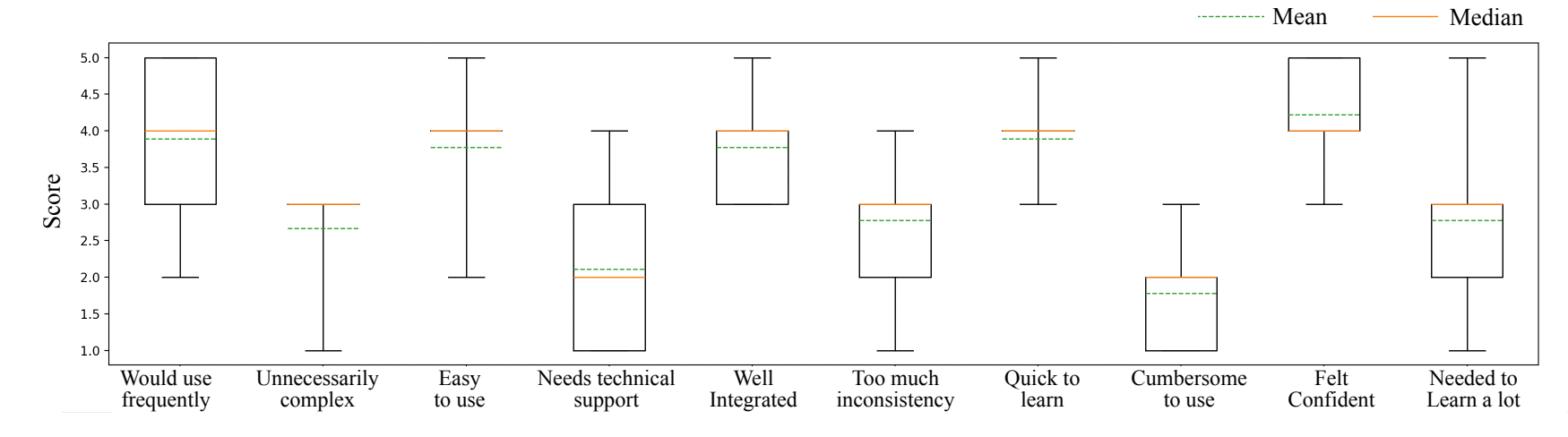}
    \caption{SUS item results, aggregated over participants ($\bullet$ Mean, \textbf{---} Median, \textcolor{susd}{\textbf{- -}} Benchmark Average, \textcolor{susu}{\textbf{$\cdot \cdot \cdot$}} Benchmark Good)}
    \Description[Box and whisker plot showing SUS results]{Results: would use frequently $3.80 \pm 1.05$, unnecessary complex $2.67 \pm 0.71$, easy to use $3.78 \pm 0.83$, would need technical support $2.11 \pm 1.27$, well integrated $3.78 \pm 0.83$, too much inconsistency $2.78 \pm 0.97$, quick to learn $3.88 \pm 0.60$, cumbersome to use $1.78 \pm 0.83$, felt confident $4.22 \pm 0.67$, needed to learn a lot $2.77 \pm 1.20$}
    \label{fig:sus}
\end{figure}

System usability was generally rated positively (see Figure~\ref{fig:sus}), with an average score of $74.9\pm12.13$.
Based on benchmarks of previously collected SUS data, this result is regarded as ``good''~\cite{bangor2009determining}.
Considering the specific questionnaire items, in relation to prior benchmarks~\cite{lewis2018item}, we note that most questions score ``above average''.
Specifically the participants were eager to use the system frequently (\textbf{Q1}, $3.80\pm1.05$), did not find it unnecessarily complex (\textbf{Q2}, $2.67\pm0.71$), and found it easy to use (\textbf{Q3}, $3.78\pm0.83$).
They did not feel the need for technical support (\textbf{Q4}, $2.11\pm1.27$), and perceived the functionalities to be well integrated (\textbf{Q5}, $3.78\pm0.83$).
Participants did not perceive the system to be cumbersome (\textbf{Q8} $1.78\pm0.83$), and they felt confident using it (\textbf{Q9}, $4.22\pm0.67$).
However, while they found the system to be quick to learn (\textbf{Q7}, $3.88\pm0.60$), 
they felt that they needed to learn a lot of things before using it (\textbf{Q10}, $2.77\pm1.20$).
Also, they perceived the system to be inconsistent (\textbf{Q6}, $2.78\pm0.97$). We suspect this is due to the navigation issues outlined above.

\section{Discussion}
\label{sec:discussion}

We discuss the key results of our study, highlighting the way we designed the system and summarizing the types of interactions that BLV participants engaged in with our LLM-enabled DATM. In addition, we address the main limitations of our methodological approach.

\subsection{System Design}
\label{sub:system_design}
The system was designed iteratively in multiple stages:
This cycle of iterations allowed to identify the use cases (\textbf{Rq1}) and to ensure that the development of new functionalities was heavily influenced by observations of BLV participants, with the functionality being iteratively debugged and improved after live participant experiments.

A key to the capability of the final system was the augmentation of the LLM with specific tool calls. In the initial iterations of the system we had expected that the LLM could handle most functionality, aside from modules such as computer vision, STT and TTS. Over time, however, the iterative prompt-building process demonstrated the inability of LLM models to reliably answer to directly user queries (\textbf{Rq2}), a problem that was addressed by refining the contextual information provided to the LLM and by incorporating multiple tool calls to ensure reliable spatial reasoning (\textbf{Rq2.1}).
While the reliance on tool calls limits the generalizability of the solution, we anticipate that the spatial reasoning abilities of LLMs will become more robust in the future, reducing or eliminating the need for tool calls.

\subsection{Observed Interactions}
\label{sub:obs_interactions}
BLV participants in the final evaluation interacted with \system{} with a combination of gestures and vocal interactions (\textbf{Rq3}). They perceived most system functionalities as useful and clear (\textbf{Rq4}), despite some initial difficulties using \system{}. Indeed, a substantial amount of training was required to accomplish this. This is partly due to the need for non-visual explanations, \textit{e.g.}, how to hold the finger to make a pointing gesture and how to operate the Bluetooth button to ask a question, which tend to require more verbalization (and time) than the types of visual explanations that sighted individuals typically incorporate (\textit{e.g.}, a photo of the desired pointing gesture).

Most participants found the pointing functionality useful and accurate, even if it sometimes triggered announcements with too much information. After training, participants were able to make the pointing gesture reliably. Problems occasionally arose in pointing gesture recognition, mostly when the pointing finger was not held sufficiently parallel to the TMAP surface, causing perspective foreshortening in the camera's view of the finger (which made finger tracking unreliable). Since conducting the study, the authors have devised a new version of the pointing gesture recognition algorithm that addresses this problem, which will be incorporated in the future.

Many participants appreciated the ability to converse with the LLM and the opportunity to ask a wide range of questions, as well as the ability to bookmark POIs for later reference. However, the over-sensitivity of the Bluetooth button used to record the voice
caused some confusion. Delays in the LLM response were also inconvenient, especially when the response was lengthy.

We observe that different participants had different ways to respond to navigation instructions, as previously observed in the literature for egocentric navigation assistance \cite{ohn2018variability}.
Indeed, participants found the fly-me-there and street-by-street navigation functions the most difficult part of the system, but most were able to master these functions after enough practice, and most acknowledged the value of receiving guidance to a POI. The challenges were caused by a few factors, including (a) confusion about the meanings of the cardinal directions (north, west, southeast, etc.) on the physical TMAP and (b) feedback delays, which forced participants to move their pointing finger slowly and often resulted in announcements that were not timely.

These challenges will be mitigated in the future, e.g., (a) the user could choose to receive directions in different formats (such as left, down, or 2 'oclock) and (b) using a faster system (see Sec.~\ref{sub:limitations})
Another factor that possibly influenced the usability of these functionalities is that their need emerged in the second preliminary study and hence, differently from the other functionalities, they have not been iteratively refined.

Overall perceptions of the system were positive, with several participants excited by the access that it provided to many forms of information, despite the limitations they noted.

\subsection{Limitations}
\label{sub:limitations}
A lengthy training session was required, which meant that the experimenters had limited time to assess participants' ability to use the system independently (to prevent the entire session from being too long). A longer assessment would have allowed us to explore how participants used the system to solve a larger variety of tasks. Moreover, our assessment consisted entirely of a set of tasks specified by the experimenters, which limits our ability to observe how participants would use the system in actual use cases.

The interface was constrained by the need to call OpenAI's GPT-4o for each query, which created unpredictable delays (from under a second to more than $10$ seconds). In the future we will explore LLM models that can be run locally on the PC, such as LLama 2, which we anticipate will solve this problem. In addition, the limited frame rate of the overall system (roughly 10 frames per second) made it difficult for \system{} to keep up with participants' moving fingers. We believe that system performance can be substantially improved by engineering the software.

At the time that we developed our prototype system, LLM models with integrated voice interactions were not yet available. As a result, we were forced to use a STT and a TTS, which resulted in an unnatural interface requiring the user to hold down the ``talk'' button while recording their question. In the future more powerful LLM models will be available with integrated voice interactions, which should allow the user to pose queries such as, `` Hey \system{}, where is the nearest Indian restaurant?''



\section{Conclusions and Future Work}
\label{sec:conclusions}
This paper describes \system{}, a novel system, that augments a tactile map with a LLM to create a conversational interface for BLV users. The system adds audio interactivity to the tactile information present in the map, enabling the user to receive immediate TTS announcements about features they point to on the map, and also to verbally ask a wide range of questions to the system.
In addition, navigation guidance functions give the user audio directions to guide them to a desired POI on the map. The system was developed in an iterative process combining feedback from BLV participants, an iterative prompt engineering process to optimize the accuracy and usefulness of the LLM responses, and experiments with BLV participants using the prototype. Tests were conducted with $10$ BLV participants using the system, demonstrating their ability to perform a set of tasks relating to a realistic scenario in which the participant is asked to plan an evening with friends in an unfamiliar urban setting. We analyze the type of questions the users are willing to ask and study how the users interact with the system, its usability, and user experience. 

In future work, we plan to exploit recent improvements to LLM models that offer integrated voice interactions. We expect that ongoing improvements in LLM spatial reasoning abilities will allow the updated MapIO system to reliably answer an even wider range of questions, with less reliance on ad hoc prompt engineering and tool calls that limit the generalizability of the current system.

\bibliographystyle{ACM-Reference-Format}
\bibliography{paper}

\end{document}